# High-speed and high-efficiency three-dimensional shape measurement based on Gray-coded light

**Zhoujie Wu, Wenbo Guo, Yueyang Li, Yihang Liu, and Qican Zhang***

Sichuan University, Department of Opto-Electronics, Chengdu, China, 610065

**Abstract:** Fringe projection profilometry has been increasingly sought and applied in dynamic three-dimensional (3D) shape measurement. In this work, a robust and high-efficiency 3D measurement based on Gray-code light is proposed. Unlike the traditional method, a novel tripartite phase unwrapping method is proposed to avoid the jump errors on the boundary of code words, which are mainly caused by the defocusing of the projector and the motion of the tested object. Subsequently, the time-overlapping coding strategy is presented to greatly increase the coding efficiency, decreasing the projected number in each group, e.g. from 7 (3 + 4) to 4 (3 + 1) for one restored 3D frame. Combination of two proposed techniques allows to reconstruct a pixel-wise and unambiguous 3D geometry of dynamic scenes with strong noise using every 4 projected patterns. The presented techniques preserve the high anti-noise ability of Gray-coded-based method while overcoming the drawbacks of jump errors and low coding efficiency. Experiments have demonstrated that the proposed method can achieve the robust and high-efficiency 3D shape measurement of high-speed dynamic scenes even polluted by strong noise.

*Qican Zhang, E-mail: zqc@scu.edu.cn

## 1 Introduction

Optical three-dimensional (3D) profilometry has been widely used for mechanical engineering, biological recognition, machine vision, intelligent manufacturing and so on.[1-3] Among all measuring methods, fringe projection profilometry (FPP) is one of research hotspots due to its high accuracy and flexibility. Recently, with rapid advancements on hardware equipment such as the high-speed camera and the digital light processing (DLP) projector,[4-6] high-speed and high-accuracy 3D shape measurement based on fringe projection has been increasingly sought by researchers.

The introduction of electronic imaging sensors based on the charge-coupled device (CCD) or complementary metal–oxide–semiconductor (CMOS) sensors revolutionized high-speed photography, enabling captured rates of up to $10^7$ frames per second (fps).[7] Recently, Gao et. al proposed a two-dimensional dynamic imaging technique, compressed ultrafast photography (CUP), which can capture non-repetitive time-evolving events at up to $10^{11}$ fps.[8] Besides, Nakagawa presented the sequentially timed all-optical mapping photography (STAMP) that can perform single-shot image acquisition equally with short frame intervals ($4.4 \times 10^{12}$ fps) and high pixel resolution ($450 \times 450$ pixels).[9] On the other hand, using the digital micro-mirror device (DMD), DLP platform (e.g., DLP Discovery, DLP Light Commander and DLP LightCrafter) can project the binary images at a fast rate (e.g. 20 kHz for DLP Discovery 4100). Combining with the binary defocusing technique,[10] high-quality sinusoidal fringe pattern can be generated and projected at a high-speed rate. Therefore, both the speeds of projection and capture are fast enough to reconstruct dynamic scenes and the remaining problem is how to find a suitable method to realize a robust and efficient high-speed 3D measurement.

In the use of different analyzed domains to retrieve phase, FPP can be categorized into the transform-based method[11-17] and the multi-frame method[18-20]. Common transform methods include Fourier transform (FTP)[11-13], windowed Fourier transform[14,15] and wavelet transform[16,17], and all of them analyze single fringe in spatial or frequency domain. FTP is the representative single-shot transform method which is well suitable for high-speed 3D shape measurement. Since pattern switching is not required for FTP method, the simpler optical projection system without the DMD or mechanical shifting units[21,22] can be designed for high-speed measurement. To our knowledge, Zhang and Su firstly applied FTP in high-speed measurement.[23] They developed a projection system with a physical grating, so, theoretically, the reconstructed rate only depends on the maximum recording speed. Based on that system, they obtained 3D shape and deformation of rotating blades using stroboscopic structured illumination[24] and achieved 3D reconstruction of the drumhead vibration at the rate of 1000 Hz[25]. To apply FTP in more dynamic scenes, Zuo et al. proposed Micro Fourier transform profilometry, which can realize an 3D acquisition rate up to 10,000 fps.[26] However, due to the limitation of the band-pass filtering, FTP method is difficult to measure objects with sharp edges, abrupt change or non-uniform reflectivity. In addition, all the transform methods have to carefully choose parameters, such as the size and the localization of the filtering window or the scale of the wavelet to obtain the high-quality retrieved phase. Consequently, for the transform methods, it is hard to realize automatic processing in complex dynamic scenes where the shape of the objects is time-varying.

Compared with the transform methods, the multi-frame methods are used more widely in optical metrology because of its high precision, low complexity and easy accomplishment under computer control. In multi-frame methods, phase extracting is achieved in temporal domain. A series of fringe patterns are projected onto the tested object surfaces at different instants and the phase for each given pixel of captured images can be independently calculated by the intensity values at that point over time. Hence, the multi-frame methods are insensitive to the varying reflectivity. Among the multi-frame methods, phase-shifting profilometry (PSP)[19,20,27] is most widely used.

In addition, the use of arctangent function caused the phase ambiguity in both FTP and PSP. So, temporal phase unwrapping (TPU) algorithms[28] are applied to eliminate the phase ambiguity and measure isolated or steep objects. Typical TPU approaches applied in dynamic 3D shape measurement can be classified into two categories: multi-frequency (or multi-wavelength) approaches[28-31] and Gray-coded-based approaches[32,33]. Multi-frequency approaches eliminate the phase ambiguity by projecting other groups of shifting fringe patterns. According to the different principles to eliminate the phase ambiguity, multi-frequency approaches can be further classified into hierarchical approach[28,29], heterodyne approach[30,31] and number-theoretical approach[34,35]. In high-speed 3D shape measurement, Wang and Zhang applied multi-frequency (heterodyne, 3 × 3) phase-shifting technique with optimal pulse width modulation to develop a 556 Hz measuring system.[36] However, it cost 9 patterns to reconstruct one result, which is relatively inefficient. Therefore, they used two-frequency (hierarchical, 3 + 3) technique to measure the shape of a live rabbit heart.[37] Then, Zuo et al. proposed high-speed measuring method using bi-frequency (number-theoretical, 3 + 2) tripolar pulse-width-modulation technique to achieve 1250 Hz 3D measurement.[38] All the traditional two-frequency TPU approaches are easy to suffer the phase-unwrapping errors when the high frequency is chosen far higher than the low frequency.[39] To improve the period number of the high-frequency fringe pattern, Yin et al. applied the depth constraint in number-theoretical approach and improved the measuring accuracy at the cost of a limited measuring depth.[40]

Gray-coded-based approaches eliminate the phase ambiguity by projecting a series of binary Gray-coded patterns, and N patterns can uniquely label $2^N$ stripe periods. Gray-coded-based TPU approaches have been widely used in 3D shape measurement for static scenes owing to its robustness and anti-noise ability. However, it is still challenging for this method to realize a high-speed shape measurement. Two essential problems, jump errors and low coding efficiency, are obliged to be solved. Because of the object's motion and the projector's defocus, jump errors occur easily on the boundaries of the adjacent Gray-coded words. To overcome this drawback, Wang et al. combined the conventional spatial phase unwrapping with Gray-coded method to solve the motion-induced phase unwrapping errors.[41] This framework is great but only works well for simple and smooth objects. Laughner et al. projected additional white image with all "1" and black image with all "0" to binarize the Gray codes and restored the accurate rabbit cardiac deformation at 200 fps.[42] But 10 images are used to recover one 3D frame. Zheng et al. projected an additional binary pattern to construct the image with all "0.5" utilizing the defocus of the projector.[43] Then the image with "0.5" grayscale is used to binarize the Gray codes. However, the binarization reduced but not eliminated the jump errors and the other correction such as median filter was required to remove the remaining errors.[44] Wu et al. recoded Gray codes in temporal and spatial domains respectively, realizing the simple and robust phase unwrapping in dynamic 3D shape measurement without projecting additional patterns.[45,46] On the other hand, Gray-coded-based methods require a large number of additional patterns for phase disambiguation (e.g. at least 4 patterns are needed to label 16 stripe periods). In our previous work[45,46], every 3 phase-shifting sinusoidal patterns and 4 Gray-coded patterns (3 + 4) are obtained to restore one 3D frame. To improve the coding efficiency, Zheng et al. proposed ternary and quaternary Gray-coded-based phase unwrapping methods using binary defocusing projection, in which several gray levels instead of black and white are used to encode the fringe period.[47,48] It acquires higher coding efficiency at the cost of reducing the signal to noise ratio. So, how to realize the robust and high-coding-efficiency 3D shape measurement based on Gray-coded light in dynamic scenes is a problem needed to be concerned.

To this end, a high-speed and high-efficiency 3D shape measurement method based on Gray codes is proposed. The tripartite phase unwrapping method is proposed to avoid the jump errors on the boundary of code words, which are mainly caused by defocusing and motion. Three staggered wrapped phases with 2π/3 phase difference can be acquired just by changing the built-up sequence of captured sinusoidal patterns, and the reference wrapped phase is created utilizing existing reference plane and decoding word to divide decoding orders into the tripartition (low region, middle region and high region). Hence, decoding orders on different regions are used to unwrap corresponding wrapped phase, enabling nonuse of wrapped phase on the jump regions and the jump errors can be pre-avoided rather than post-eliminated without additional pattern projection. Subsequently, time-overlapping Gray-coded coding strategy is presented to greatly increase the coding efficiency. Owing to the proposed tripartite phase unwrapping method and high-speed hardware, each of traditional 4 Gray-coded patterns (for labeling a 16-period fringe) is projected after every three dithering sinusoidal patterns. Therefore, every Gray-coded pattern could be used four times to decrease the projected number in each group from 7 (3 + 4) to 4 (3 + 1). Combination of two proposed techniques allows to reconstruct a pixel-wise and unambiguous 3D geometry of dynamic scenes with every 4 projected patterns. The presented techniques preserve the high anti-noise ability of Gray-coded-based method while overcoming the drawbacks of jump errors and low coding efficiency, and finally achieve the robust and high-efficiency 3D measurement in high-speed dynamic scenes even polluted by strong noise.

## 2  Principle

Our high-speed measurement system includes a DLP projector and a high-speed camera as shown in Fig. 1. Firstly, the projecting speed can be greatly increased by applying binary defocusing technique[10] in which binary pseudo-sinusoidal patterns are projected by slightly defocusing the projection lens. Then, high-quality sinusoidal fringe is generated on the surface of objects and the high-speed camera captures the deformed patterns in sync with the projector. Next, fringe analysis and phase unwrapping algorithms are used to obtain the unambiguous phase. Finally, 3D shape information can be acquired through system calibration.

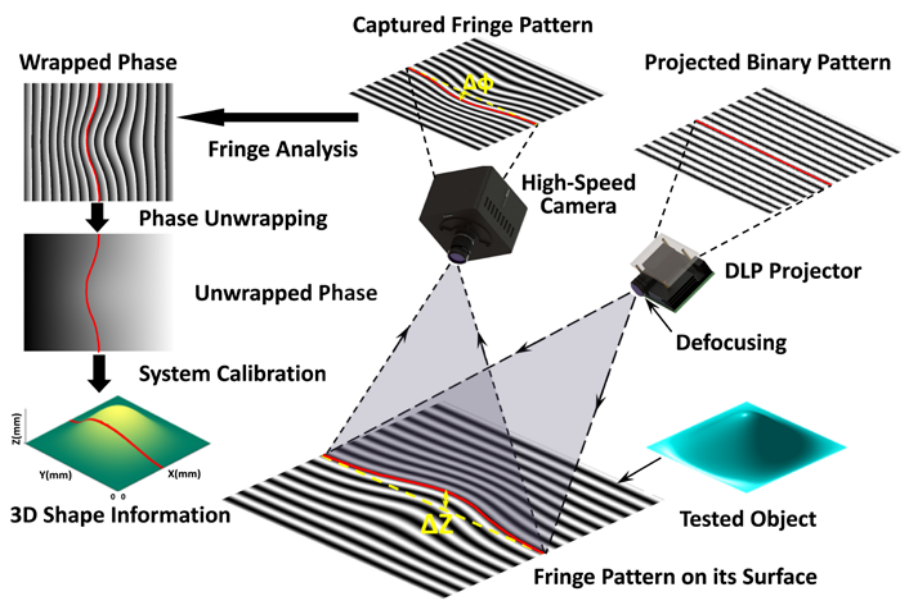

Fig. 1. Schematic diagram of the high-speed measurement system.

*2.1 Integration of Phase-shifting Algorithm with Gray-coded Method.*

Either PSP or Gray-coded method (GCM)[49] can resolve 3D geometry. Integration of PSP with GCM brings together the advantages of both strategies, i.e. the high spatial resolution of PSM, plus the unambiguity and robustness of GCM.[50] Since at least three patterns are required to extract the phase in PSP, the three-step phase-shifting algorithm is widely used to reduce the measuring time and the motion-induced errors in dynamic measurement. And three phase-shifting sinusoidal patterns can be described as:

$$I_n(x, y) = A(x, y) + B(x, y)\cos[\phi(x, y)+2\pi(n-1)/3+\phi_0], \quad n=1,2,3 \qquad (1)$$

in which, $A(x, y)$ is the intensity of the background light, $B(x, y)/A(x, y)$ is the fringe contrast and $\phi(x, y)$ is the wrapped phase of the modulated light field which can be obtained using three-step phase-shifting algorithm described in Eq. 2. In addition, $\phi_0$ is the initial phase value to match the wrapped phase with the Gray codes.

$$\phi(x, y) = \tan^{-1} \frac{\sqrt{3}(I_1(x, y) - I_3(x, y))}{2I_2(x, y) - I_1(x, y) - I_3(x, y)} \qquad (2)$$

The periodic nature of the sinusoidal patterns introduces phase ambiguity; hence, Gray-coded patterns are projected to label the fringe order and eventually eliminate phase ambiguity. In the traditional Gray-code based TPU method, $N$ Gray-coded patterns can label maximum $2^N$ fringe orders. And the phase order $k(x, y)$ can be calculated using following equations:

$$V(x, y) = \sum_{i=1}^{N} GC_i(x, y) * 2^{(N-i)}, \qquad (3)$$

$$k(x, y) = i(V(x, y)), \qquad (4)$$

in which, $GC_i(x, y)$ denotes the $i_{th}$ binarized Gray-coded pattern, $V(x, y)$ is the decoding decimal number and the function $i(\cdot)$ is used to look up the known unique relationship between $V(x, y)$ and the $k(x, y)$. For convenience of describing, the stripe period is assumed as 4, so 2 Gray-coded patterns are required to be projected to uniquely label different regions as shown in Fig. 2 (a).

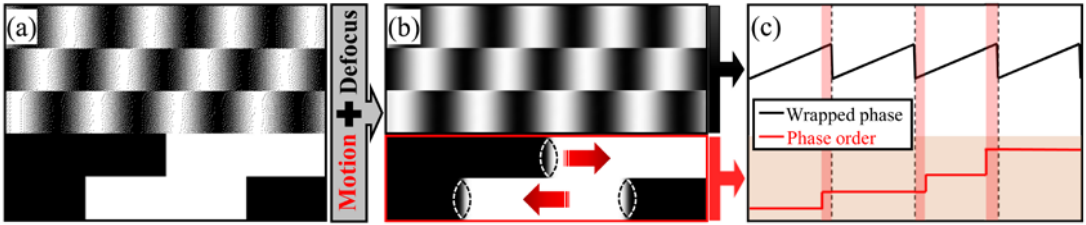

Fig. 2. Cause of the jump errors of the Gray-code-based TPU method in dynamic measurement. (a) Projected binary patterns. (b) Acquired grey-scale patterns after defocus and motion. (c) Mismatch between the wrapped phase and the phase order.

To achieve high-speed projection, the dithering technique[51] is applied to produce high-quality sinusoidal fringes by defocusing the projector as shown in Fig. 2 (a). However, the boundaries of black and white value conversion of Gray-coded patterns are also blurred as show in Fig. 2(b), resulting in inaccurate binarization. Besides, the motion of the objects also brings in the shift among patterns. These two main factors will cause the mismatch between the wrapped phase and the phase order as shown in Fig. 2(c). Therefore, jump errors easily occur on the boundaries of adjacent Gray-coded words. Accordance to the Gray-coded design manner, the Hamming distance of two adjacent codewords is 1, which makes the wrong orders on the boudaries only have one-order jump errors. It means that the phase order map still remains stair-stepping as shown in Fig. 2(c), which is the precondition of the proposed phase unwrapping method described in the next subsection.

*2.2 Tripartite Phase Unwrapping Method.*

The jump errors on the boundaries will cause $2\pi$ phase errors in the phase unwrapping process. Most previous methods overcome this drawback through filtering[44] or monotonicity detection[52] to detect and correct phase errors. These methods are post-eliminated means and perform well in static scenes. But it is hard to apply these methods in dynamic measurement because the width of the error regions might be larger in dynamic scenes, causing difficulty to distinguish the jump errors from the abrupt change of the object. Our previous works[45,46] used pre-avoided methods and achieved robust measurement for complex dynamic scenes with abrupt changes by utilizing Gray codes in adjacent projected sequences, which confined the further promotion of the Gray-coded coding efficiency in high-speed measurement. Therefore, in this subsection, a novel pre-avoided method named tripartite phase unwrapping (Tri-PU) method is proposed.

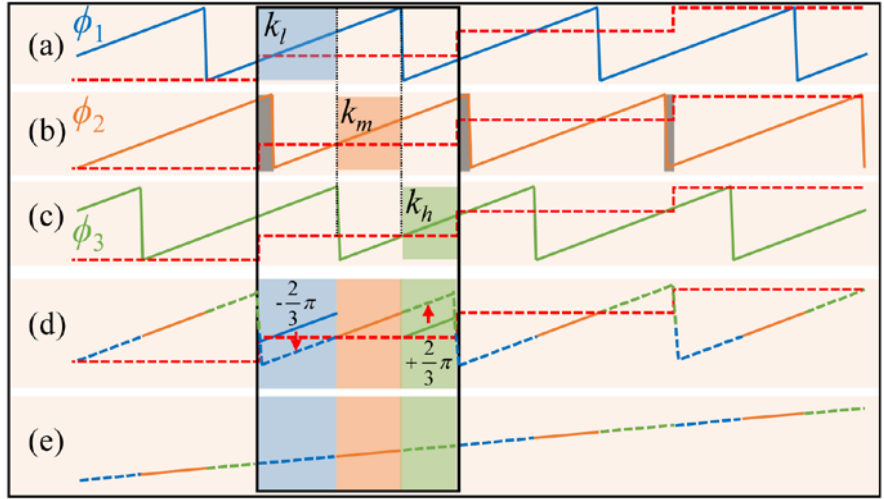

Fig. 3. Schematic diagram of the tripartite phase unwrapping method. (a) Wrapped phase $\phi_1$ calculated from [$I_2,I_3,I_1$]. (b) Wrapped phase $\phi_2$ calculated from [$I_1,I_2,I_3$]. (c) Wrapped phase $\phi_3$ calculated from [$I_3,I_1,I_2$]. (d) Phase compensation for $\phi_1$ and $\phi_3$. (e) Unwrapped phase.

Several staggered wrapped phases with fixed phase difference can be acquired just by changing the built-up sequence of captured sinusoidal patterns.[53] So as shown in Figs. 3(a)-3(c), three wrapped phases $\phi_1$, $\phi_2$ and $\phi_3$ can be calculated by applying different pattern sequences [$I_2$, $I_3$, $I_1$], [$I_1$, $I_2$, $I_3$] and [$I_3$, $I_1$, $I_2$] in Eq. 2, respectively. In fact, the shift of the pattern sequences only changes the initial phase vaule with $2\pi/3$, causing the staggered discontinuity in three different wrapped phase. Hence, if the phase order $k$ can also be divided into three different regions $k_l$ (low part of $k$), $k_m$ (middle part of $k$) and $k_h$ (high part of $k$) as shown in the red rectangle of Fig. 3, the different parts of $k$ can be used to unwrap the corresponding parts in $\phi_1$, $\phi_2$ and $\phi_3$, respectively. In this way, the boundaries of $k$ (grey shaded area in Fig.3 (b)) which suffer the jump errors are used to unwrap the continuous parts of $\phi_1$ and $\phi_3$ to effectively avoid the jump errors. As shown in Fig. 3, three parts of the $k$ are colored in the same with the corresponding wrapped phases. So, as shown in Fig. 3 (e), the phase ambiguity can be eliminated using Eq. 5, once the initial phase difference $2\pi/3$ is compensated.

The tripartite phase unwrapping strategy works well when the mismatch between the wrapped phase and the phase order is no more than 2/3 period, which is usually easy to be guaranteed in dynamic measurement.

$$\Phi(x, y) = \begin{cases} \phi_1(x, y) + 2\pi k_l(x, y) - 2\pi/3, & k \in k_l \\ \phi_2(x, y) + 2\pi k_m(x, y), & k \in k_m \\ \phi_3(x, y) + 2\pi k_h(x, y) + 2\pi/3, & k \in k_h \end{cases} \quad (5)$$

It should be mentioned that the reason of the jump errors is the edges mismatch of the wrapped phase and the phase order, and the common approach to solve this problem relies on the edge of the wrapped phase and corrects the edge of the phase order. However, in this work, the edges of the phase order are regarded as reliable and the relatively unreliable discontinuities of the wrapped phase are avoided by shifting the position of the discontinuities. Therefore, in the next subsection, the region division of the phase order cannot rely on the edge of the wrapped phase $\phi_1$, $\phi_2$ and $\phi_3$.

*2.3 Regional Division Using Reference Wrapped Phase.*

To apply the proposed Tri-PU method, the phase order is acquired to be divided. Different from the wrapped phase, phase order doesn't have monotonicity but stays unique in every single order. The assisted information is acquired to divide every phase order into three parts. Therefore, we propose to use the artificially created reference wrapped phase $\phi_{ref}$ to divide each class of the phase order into tripartition.

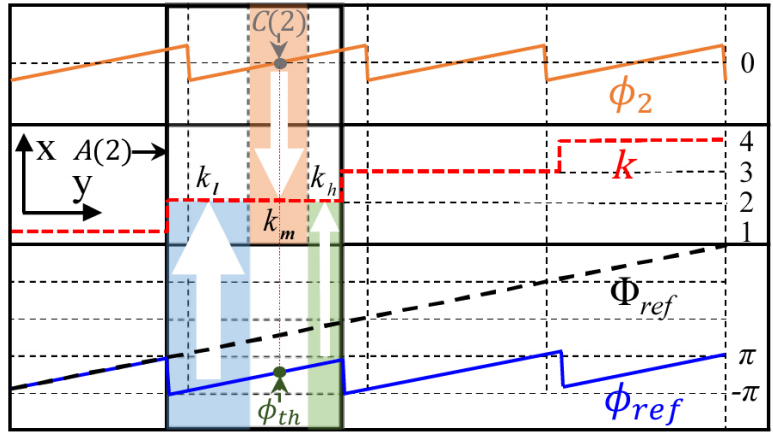

Fig. 4. Schematic diagram of the regional division using reference wrapped phase.

In fact, the mismatch between the wrapped phase and the phase order only occurs on the two boundaries (left and right) of each *k* as shown in Fig. 4, so the middle region $k_m$ can be easily found by the phase value of $\phi_2$ using following equation:

$$k_m(x, y) = k(x, y), \text{ where } |\phi_2(x, y)| < \pi/3. \tag{6}$$

Hence, the problem is simplified to distinguish the left low region and right high region of *k*. The existing wrapped phases ($\phi_1$, $\phi_2$ and $\phi_3$) lose the monotonicity in the edge regions of *k*. Therefore, an artificially created reference wrapped phase $\phi_{ref}$ is created:

$$\phi_{ref}(x, y) = \Phi_{ref}(x, y) - 2\pi k(x, y), \tag{7}$$

in which, $\Phi_{ref}$ is the absolute phase of a measured reference plane which is already determined once the measuring system has been set up and has been obtained in the calibration process and $\phi_{ref}$ is a created wrapped phase that has the same discontinuous positions with *k* as shown in Fig. 4. In this way, a phase map with monotonicity in every single phase order can be acquired to assist to distinguish the remaining two regions of *k* in each order. Firstly, we define region $A(i)$ ($i=1,2,3…2^N$) for each fringe order where $k(x, y)=i$. Since $\Phi_{ref}$ is the absolute phase of measured reference plane rather than the object, $\phi_{ref}$ changes out of the range of $(-\pi, \pi]$, causing that the critical threshold $\phi_{th}$ of $\phi_{ref}$ in $A(i)$ is no longer the zero and varies in different periods. Therefore, $\phi_{th}$ in $A(i)$ is determined by finding the critical line $C(i)$ where $\phi_2$ has the minimum absolute value along axis $y$ in $A(i)$ using Eq. (8).

$$\phi_{th}(x) = \phi_{ref}(x, y), (x, y) \in C(i) \tag{8}$$

So, phase values of $\phi_{ref}$ on the $C(i)$ can be used as $\phi_{th}$ to determine the lower region $k_l$ and the higher region $k_h$ using Eq. (9).

$$\begin{aligned} k_l(x, y) &= k(x, y), \ where \ \phi_{ref}(x, y) < \phi_{th}(x), \ (x, y) \in A(i) \ and \ (x, y) \notin k_m \\ k_h(x, y) &= k(x, y), \ where \ \phi_{ref}(x, y) > \phi_{th}(x), \ (x, y) \in A(i) \ and \ (x, y) \notin k_m \end{aligned} \tag{9}$$

In the most cases, the proposed regional division algorithm can perform well, but caused by the shadow and the edge of the tested objects, the incorrect regional division will occur in few specific areas only with fringe information in narrow width. Thus, the correction algorithm is presented to improve the robust of the proposed method. The detailed information is discussed in appendix.

*2.4 Time-overlapping Gray-coded Coding Strategy.*

The proposed Tri-PU method can solve the jump errors problem in the Gray-coded-based method without using additional patterns. In this work, the number of projected Gray codes is assigned as 4 to compromise the measuring accuracy and speed. However, 7 patterns (3 phase-shifting patterns and 4 Gray-coded patterns) are still required to be projected in every sequence, which is relatively inefficient compared with other methods in high-speed measurement. Therefore, the time-overlapping Gray-coded coding strategy is proposed in this subsection to significantly increase the coding efficiency of the Gray-coded-based method in high-speed measurement.

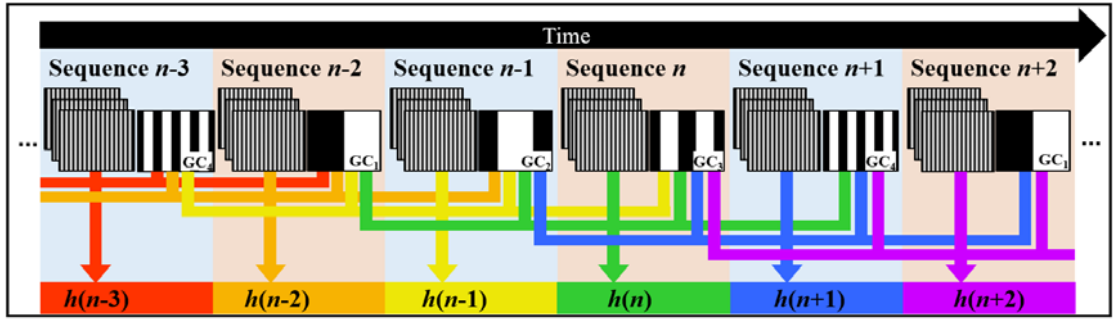

Fig. 5. Time-overlapping Gray-coded coding strategy.

As shown in Fig. 5, each of traditional 4 Gray coded patterns are projected after every three dithering sinusoidal patterns rather than continuously projecting 4 patterns in every sequence. And for one group of sinusoidal patterns, the adjacent 4 Gray-coded patterns are used to eliminate the phase ambiguity as shown in Fig. 5 (used patterns in one sequence are connected in a same color). Therefore, every Gray-coded pattern could be used four times to greatly decrease the projected number in every projecting group from 7 (3 + 4) to 4 (3 + 1). This time-overlapping strategy is suitable to be used in our work to improve coding efficiency owning to the high-speed hardware and the proposed Tri-PU method. In high-speed 3D measurement, the projecting and capturing speeds are commonly set much higher (over several kHz) than that of the measured objects to ignore the motion-induced phase-shifting errors using the high-speed hardware. Hence, the motion-induce disparity of the adjacent several patterns is small and the discontinuous projection of the Gray-coded patterns can be ignored compared with the threshold 2/3 period of the Tri-PU method. In addition, in the Tri-PU method, the motion-induced mismatch between the sequential captured patterns will not influence the phase unwrapping successful rate when the mismatch length between the wrapped phase and the phase order is less than 2/3 period. Because the motion-induced mismatch only influences the phase unwrapping successful rate on the discontinuous parts of the wrapped phase. But in the Tri-PU method, only the middle regions of the phase orders are used. Therefore, the time-overlapping strategy is suitable to be used in our work to improve coding efficiency.

*2.5 System Calibration.*

In order to obtain the 3D surface information of the object, the absolute phase is needed to be converted to the height using the phase-to-height algorithm.[54] Equation (10) can be used to reconstruct the height:

$$\frac{1}{h(x,y,n)} = u(x,y) + \frac{v(x,y)}{\Delta\Phi(x,y,n)} + \frac{w(x,y)}{\Delta\Phi^2(x,y,n)}, \quad (10)$$

in which, $\Delta\Phi(x, y, n)$ is the absolute phase value of the measured object, relative to the reference plane. Four planes which have known height distributions are measured. Then the 3 unknown parameters $u(x, y)$, $v(x, y)$ and $w(x, y)$ can be calculated and saved as system parameters for the future phase-to-height mapping.

The camera calibration technique proposed by Zhang[55] is widely adopted and used in computer vision, so it is implemented to calibrate the camera in our developed system.

## 3 Experiments and results

Experiments have been conducted to test the performance of our proposed method. A measuring system was developed, including a DLP projector (LightCrafter 4500) and a high-speed CCD camera (Photron FASTCAM Mini UX100). The projector resolution is $912 \times 1140$ pixels and the lens assembled to the camera has a focal length of 16mm and an aperture of f/1.4. In all our experiments, the image refreshing rate of the projector was set at 2170Hz and the period number of the projected sinusoidal fringes is 16. And $912 \times 1120$ pixels of the projector are used to generate the projected fringe patterns whose period is 70 pixels. The camera resolution was set at $1280 \times 1000$ pixels and the camera was synchronized by the trigger signal of the projector.

*3.1 Measurement on the Static Scene.*

In the first experiment, a cooling fan for central processing unit and a portrait sculpture were measured using the proposed Tri-PU method and the traditional phase unwrapping (Tra-PU) method to demonstrate the validity of the Tri-PU method and compare the performance of two methods. To explain the whole procedure of the Tri-PU method clearly, the whole framework of this method is illustrated in Fig. 6 (a). Firstly, 3 deformed phase-shifting patterns are acquired to calculate three staggered wrapped phases $\phi_1$, $\phi_2$ and $\phi_3$ with $2\pi/3$ phase difference just by changing their built-up sequence and 4 Gray-coded patterns are obtained to calculate the phase order $k$. Then, the unwrapping phase $\Phi_{ref}$ of the reference plane and $k$ are used to create the reference wrapped phase $\phi_{ref}$ using Eq. (7). Next, three parts $k_h$, $k_m$ and $k_l$ can be accurately divided and labeled in every phase order using $\phi_{ref}$ and $\phi_2$. After regional division, three wrapped phases ($\phi_1$, $\phi_2$ and $\phi_3$), three divided regions ($k_h$, $k_m$ and $k_l$) and phase order $k$ are used to eliminate the phase ambiguity using the proposed Tri-PU method; besides, $\phi_2$ and $k$ are also used to unwrap the wrapped phase using the Tra-PU method. Finally, two unambiguous phases are converted to the 3D results by the system calibration. In addition, the line profiles (labeled in red in the texture map) of the key data are shown in Fig. 6(b).

The experimental results show the proposed Tri-PU method can well avoid the jump errors in the traditional Gray-coded-based method without using additional projected patterns. In the specific areas like shadow and edge areas, the correction algorithm in Tri-PU method works well as shown in the subfigure of Fig. 6(a).

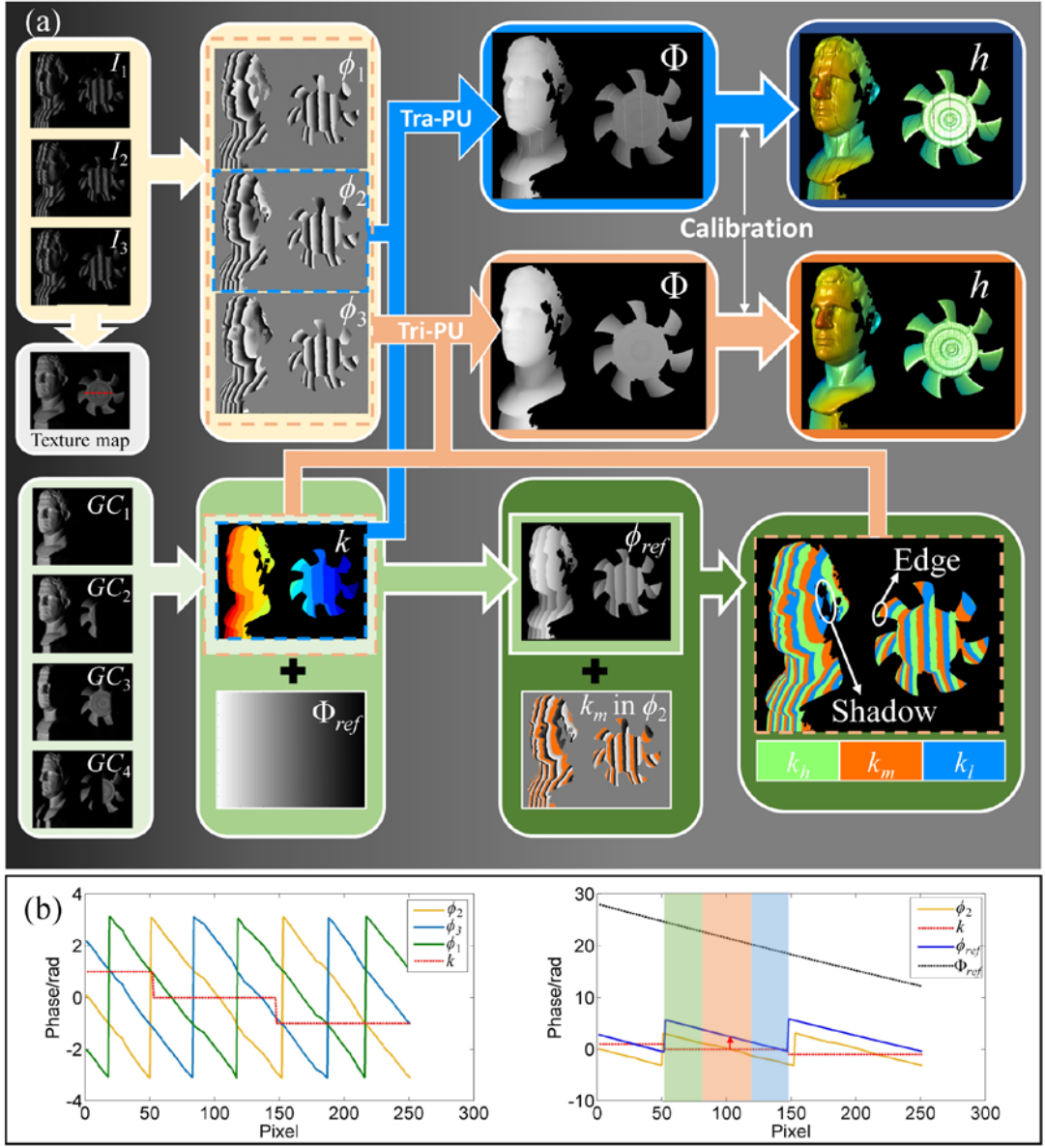

Fig. 6. Framework of the proposed method. (a) Procedure of the proposed method. (b) Line profiles (labeled in red in the texture map) of the key data in (a).

*3.2 Accuracy Analysis.*

To quantitatively evaluate the accuracy of the proposed method, a standard ceramic ball and a designed step-shaped workpiece were measured as shown in Fig. 7(a). The radius of the ceramic ball is 12.6994 millimeter (mm) which is measured by the coordinate measurement machine and the step height of the workpiece is designed to be 30 mm and the machining error is less than 30 microns. One of the deformed fringe patterns and the divided regions are shown in Fig. 7(b) and 7(c). Then, the 3D result is reconstructed using the presented method as shown in Fig. 7(d). Firstly, the flatness of the measured four planes in the workpiece was evaluated. Four fitting planes on the restored result were treated as the ground truth and the error distribution of each step is shown in Fig. 7(e). The root-mean-square (RMS) errors of four measured planes are 0.0999 mm (Plane 1), 0.0685 mm (Plane 2), 0.0594 mm (Plane 3) and 0.0762 mm (Plane 4). And because the best defocusing level is in the middle region of the calibrated volume, the flatness of the planes in the middle region is better than that in both ends. Then, the measured accuracy in Z axis was evaluated. The fitting plane of the Plane 1 was treated as the base plane and the height difference of the step is shown in Fig. 7(f). The mean of height on each plane is 0.0002 mm (Plane 1), 30.0054 mm (Plane 2), 59.9867 mm (Plane 3) and 89.9003 mm (Plane 4). Finally, the measured accuracy in the whole volume was evaluated by measuring the standard ceramic ball. As shown in Fig. 7(g), the sphere fitting on the reconstructed 3D geometry of the ceramic ball is performed and the measured radius of the ball is 12.7083 mm and the RMS error of the ball is 0.0767 mm. The error distribution of the ball is shown in Fig. 7(h). These results show that the presented system can achieve a measurement accuracy better than 0.1 mm in the measurement volume size of 180 mm $\times$ 250 mm $\times$ 90 mm.

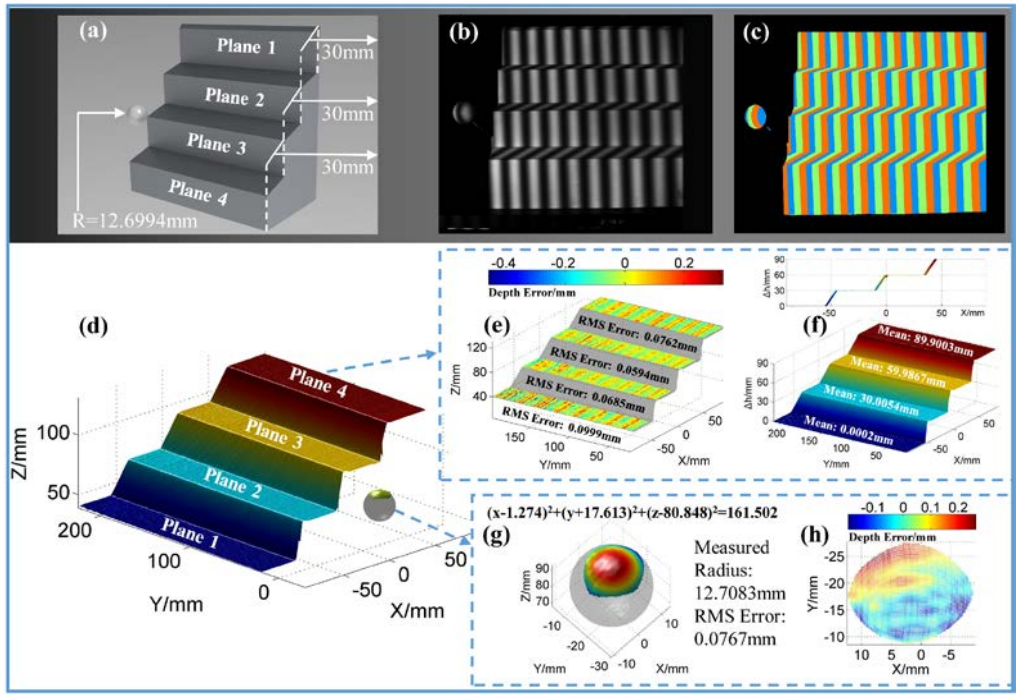

Fig. 7. Accuracy analysis of the proposed method. (a) Design drawing of the measured standard pieces. (b) Captured deformed fringe pattern. (c) Divided tripartite regions. (d) Reconstructed result. (e) Flatness error distribution. (f) Height difference of the steps. (g) Measured result and fitting sphere of the standard ball. (h) Error distribution of the standard ball.

*3.3 Measurement on the Complex Dynamic Scenes with Low SNR.*

*3.3.1 Collapsing Building Blocks.*

In this experiment, the dynamic process during which building blocks are teared down by a hand was measured to verify the performance of the proposed method in the complex dynamic scene with low signal-to-noise ratio (SNR). Without spraying the high reflectivity paint on the object's surface, building blocks have different reflectivity, texture and lots of scratches on its surface as shown in Fig. 8(e). So, it is a scene with low SNR as shown in Figs. 8(a)-8(d) for fringe projection technique. Firstly, a contrast experiment was performed to compare the anti-noise ability of the proposed method, the two-frequency method[28] and the two-wavelength method[30]. Three groups of three-step dithering phase-shifting patterns with different frequencies ($f_l$=1, $f_m$=15 and $f_h$=16) and four Gray-coded patterns are projected and captured as shown in Figs. 8(a)-8(d). Then, three height distributions are reconstructed using the two-frequency, the two-wavelength and the proposed methods respectively as shown in Figs. 8(f)-8(h). And data indicates that most phase unwrapping errors (error rate: 2.02%, which is the ratio between the numbers of error points and valid points) occurred in the result using two-wavelength method, some errors (error rate: 1.46%) occurred in the two-frequency method and none errors occurred in the proposed method. This experiment shows the proposed method has higher anti-noise ability than two-frequency and two-wavelength methods using the binary defocusing technique. In the two-frequency method, the optimal defocused degree cannot be ensured for the unit-frequency sinusoidal pattern because high-frequency sinusoidal patterns are optimally defocused to pursue higher measuring accuracy, and hence the phase quality of the unit-frequency fringe is relatively low. And in the two-wavelength method, the heterodyne algorithm increases the unambiguous measurement range by sacrificing its SNR, so this method is sensitive to the noise. Error rates increase with the ratio between two fringe frequencies in two-frequency and two-wavelength methods.[39] But in the proposed method, the error rate is independent of the fringe frequency. Because Gray codes which only have two gray scales are used to eliminate phase unambiguity and only the Gray codes are used in the decoding process. In addition, the presented tripartite phase unwrapping method is used to avoid errors at the transition area between black and white codes, so the proposed method can restore the low-SNR scene without phase unwrapping errors. This comparative experiment shows the proposed method has the better anti-noise ability and robustness than two-frequency and two-wavelength methods in the scenes with low SNR on the condition of using binary defocusing technique for high-speed measurement.

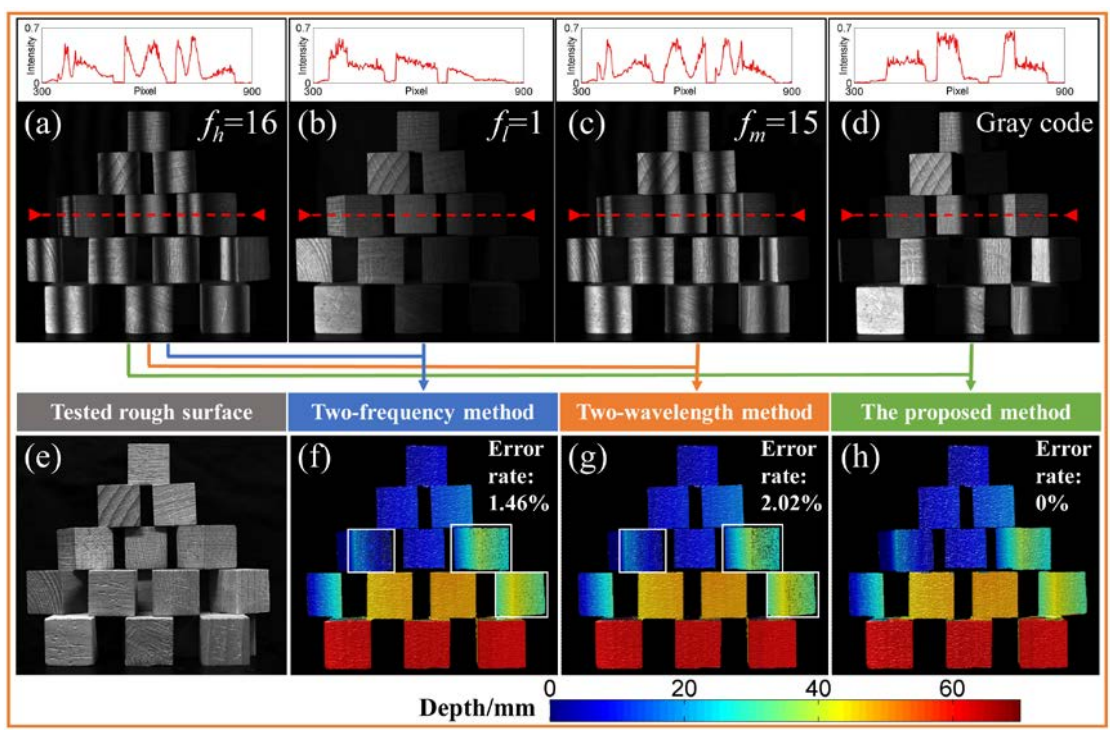

Fig. 8. Comparative experiments on the anti-noise ability. (a)-(c) Captured deformed fringe images with different frequencies ($f_h$=16, $f_l$=1 and $f_m$=15) and the intensity in line 480 of the corresponding images. (d) Captured deformed Gray-coded image and the intensity in line 480. (e) Texture map of the blocks. (f)-(h) Reconstructed results using the two-frequency, two-wavelength and proposed method respectively.

After the proof of the high anti-noise ability of the proposed method, the proposed method was applied to measure the dynamic collapsing process of this high-noise scene. Three-layer building blocks were teared down by a hand. As shown in Fig. 9(a), the proposed time-overlapping Gray-coded coding strategy is used to greatly improve the coding efficiency. Therefore, the proposed method allows to reconstruct a 3D geometry of dynamic scenes with every four projected patterns. And the texture map (average intensity of the three phase-shifting fringe images) and corresponding high-quality results at different moments are shown in Figs.

9(b) and 9(c) and Video 1. The reconstructed rate is 2170/4=542 fps and the replayed rate is 30 fps. Experimental results show that the proposed method can achieve robust measurement in the complex and dynamic scene, while guaranteeing the high coding efficiency.

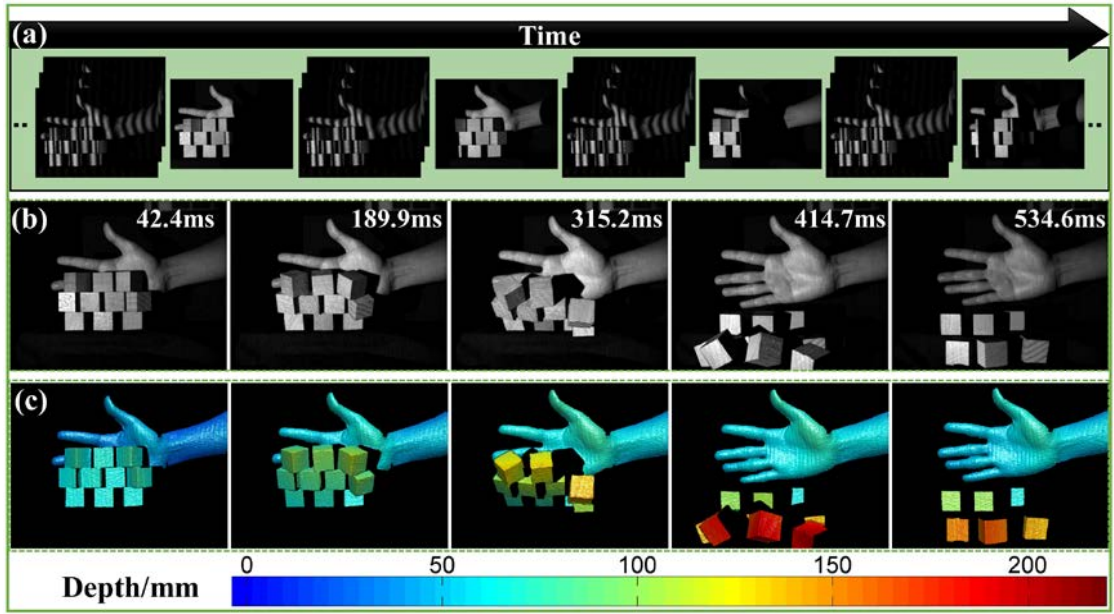

Fig. 9. Measurement on the dynamic scene of collapsing building blocks. (a) Captured pattern sequences. (b) Representative collapsing scenes. (c) Corresponding 3D frames (Video 1).

*3.3.2 Rotating Fan Blades.*

In the last experiment, the rotating fan blades were measured to further validate the high robust and anti-noise ability. With the dense speckle spraying on the blade surface, strong noise occurs in the captured images as shown in Fig. 10(b). Hence, the acquired fringe image also has low SNR as shown in Fig. 10(c). The rotation of the blades was driven by the wind from the opposite fan, and thus blades rotated in a counterclockwise direction. The reconstructed 3D frame at the time (T=274.7 ms) is shown in Fig. 10(a). And 5 profiles in the white dashed line of Fig. 10(a) at time intervals of 31.3 ms are given in Fig. 10(d), which shows the height varied process when a piece of blade rotates across one line. Furthermore, another four results with every revolution of one quarter turn are given to evaluate the rotating speed as shown in Figs.

10(e)-10(h). And the time intervals (265.5 ms, 243.3 ms and 206.4 ms) decrease progressively, which indicates the fan is in the speed-up phase and the average rotation speed is about 64 rotations per minute in this speed-up process. The corresponding 3D video is shown in Video 2. The reconstructed rate is 2170/4=542 fps and the replayed rate is 60 fps.

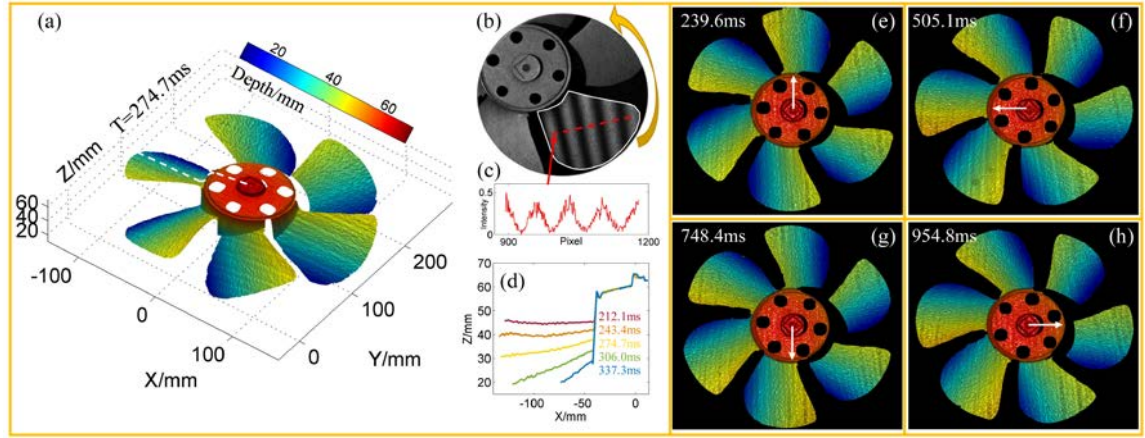

Fig. 10. Measurement on the dynamic scene of rotating fan blades. (a) Reconstructed result at the time T=274.7 ms. (b) Captured image with low SNR. (c) Intensity distribution of the red dashed line in (b). (d) Five line profiles of the white dashed line in (a) at the time intervals of 31.3 ms. (e)-(h) Four results with the interval of a quarter turn (Video 2).

## 4 Discussion

Our proposed method has the following advantages compared with other high-speed 3D measurement techniques.

The proposed method has high robustness and anti-noise ability in phase unwrapping. Compared with the two-frequency or two-wavelength method, the Gray-code based method only projects high-frequency phase-shifting patterns. Hence, the optimum defocusing degree in binary dithering technique can be guaranteed. And Gray-coded patterns which only have two gray scales are used to eliminate phase unambiguity and only the Gray-coded patterns are used in the decoding process. So, the phase unwrapping error rate is independent of the frequency of sinusoidal fringes and the noise will not be amplified in phase unwrapping process. Based on these merits, the Gray-code based method has better anti-noise ability compared with the two-frequency and two-wavelength methods in high-speed measurement. In addition, the presented Tri-PU method is used to avoid errors at the transition area between black and white codes and guarantee high robustness. So, combining the Gray-code based method and the Tri-PU method, the proposed method can realize the high robust and anti-noise phase unwrapping in dynamic scenes. It should be mentioned that the proposed Tri-PU method can be also applied in other fringe projection techniques to solve the problem of jump errors.

The proposed method has high-efficiency coding strategy in high-speed dynamic measurement. Low coding efficiency is the primary drawback of the Gray-code-based method in high-speed measurement. But owning to the high-speed hardware and the proposed Tri-PU method, the proposed time-overlapping strategy has been used to greatly improve coding efficiency. The number of projected patterns is reduced from 7 to 4 in every sequence and a pixel-wise and unambiguous 3D geometry of dynamic scenes can be restored with every 4 projected patterns. Compared with the widely-used two-frequency and two-wavelength methods, the separation of the Gray codes will not decrease the phase unwrapping successful rate, so this strategy is suitable to be used to improve coding efficiency in dynamic measurement.

The proposed method can realize robust and high-efficiency 3D shape measurement in complex dynamic scenes with low SNR. With the proposed Tri-PU method and time-overlapping strategy, robust and high-efficiency 3D shape measurement can be achieved in complex dynamic scenes with low SNR. The proposed method perseveres the high anti-noise ability of the traditional Gray-code based method while overcoming the drawbacks of the jump errors in edges and low coding efficiency.

**5. Conclusion**

In this study, high-speed and high-efficiency 3D measurement based on Gray-coded light has been proposed. The Tri-PU method is proposed to avoid the jump errors on the boundary of code words, which are mainly caused by defocusing and motion. Three staggered wrapped phases with $2\pi/3$ phase difference can be acquired just by changing the built-up sequence of captured sinusoidal patterns, and the reference wrapped phase is created utilizing existing reference plane and decoding word to divide each class of the phase order into the tripartition. And then decoding orders on different regions are used to unwrap corresponding wrapped phase, enabling the middle part of the wrapped phase on the order jump regions. Hence, the jump errors can be pre-avoided rather than post-eliminated without additional pattern projection. Subsequently, time-overlapping coding strategy is presented to greatly increase the coding efficiency, decreasing the projected number in each group from 7 to 4. Combination of two proposed techniques allows to reconstruct a pixel-wise and unambiguous 3D geometry of dynamic scenes with strong noise using every 4 projected patterns. The presented techniques preserve the high anti-noise ability of Gray-coded-based method and overcome the drawback of jump errors and low coding efficiency. Experiments has demonstrated that the proposed method can achieve the high-robust and high-efficiency 3D measurement of high-speed dynamic scenes with low SNR at the rate of 542 fps.

*Funding*

National Natural Science Foundation of China (61675141).

*Acknowledgments*

The authors would like to express sincere gratitude to Prof. Xianyu Su for his encouragement and helpful discussion.

*Appendix*

In actual measurement, the phase information in the whole period cannot be completely acquired in the shadow or the edge area. Once these areas exceed the area where $|\phi_2(x, y)| < \pi/3$, the critical line $C(i)$ may occur on the edge of the each phase order $E(i)$ as shown in Fig. 11 (a), which is labeled in red dot. And it will cause the incorrect regional division. Thus, the correction algorithm is required, in which the points of $C(i)$ which appear in $E(i)$ will be replaced with the point where $|\phi_2(x, y)|$ reach the local minimum value out of $E(i)$, which is labeled in yellow dot.

To clearly illustrate the whole procedure of the regional division algorithm with correction algorithm, a flowchart is plotted as shown in Fig. 11(b).

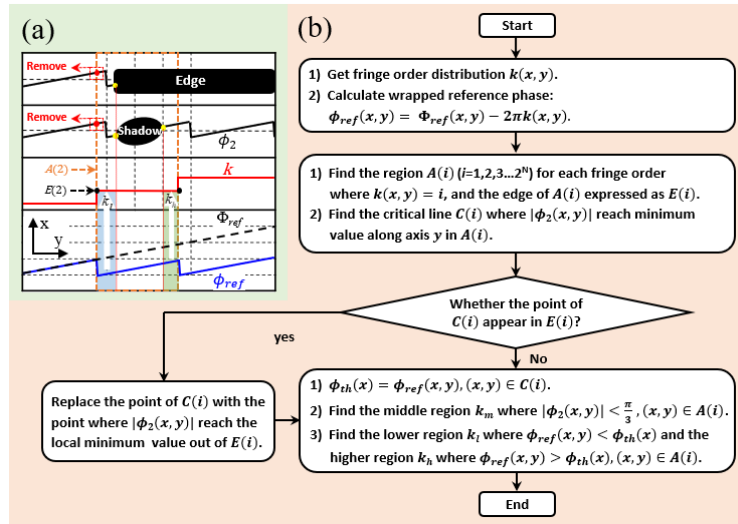

Fig. 11. Schematic diagram of the correction algorithm. (a) Schematic diagram of the errors occurred on the edge or shade regions. (b) Flowchart of the whole regional division algorithm.

As shown in Fig. 12, some errors occurred in the edge of rotating fan can be avoided using the correction algorithm. And because the errors occur in the edge or shade region with only several pixels, the median filter or erosion operation can also be applied to directly remove these errors at the cost of accuracy of some pixels.

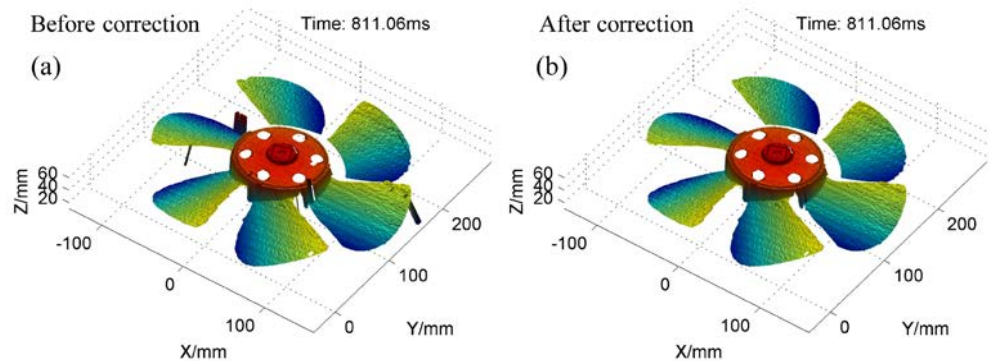

Fig. 12. Measurement results before and after correction. (a) Measurement results before correction. (b) Measurement results after correction.

**Zhoujie Wu** received his bachelor's degree in electronic science & technology from Sichuan University in 2016. He is currently a second-year PhD student at 3D sensing and machine vision Laboratory in Sichuan University. His research interest is high-speed optical three-dimensional measurement. He is a member of SPIE and was the president of the SPIE Student Club of Sichuan University from 2017 to 2018.

**Wenbo Guo** received his bachelor's degree in electronic science & technology from Sichuan University in 2017. He is currently a first-year PhD student at 3D sensing and machine vision

Laboratory in Sichuan University. His research interest is real-time optical three-dimensional measurement. He is a member of SPIE.

**Yueyang Li** received his bachelor's degree in electronic science & technology from Sichuan University in 2018. He is now a postgraduate student at 3D sensing and machine Laboratory in Sichuan University and a member of SPIE. He is mainly engaged in the study of fringe projection technology.

**Yihang Liu** is a second-year PhD student in 3D sensing and machine vision Laboratory at Sichuan University. Her research interest is high-speed 3D measurement. She is a member of SPIE and was the president of the SPIE Sichuan university chapter from 2018 to 2019.

**Qican Zhang** received his Ph.D. degree in optical engineering from Sichuan University in 2005. He is now a Professor in the Department of Electronics and Information Engineering at Sichuan University, China. His major research interest is three-dimensional (3D) optical metrology technique for dynamic object. He has published over 150 articles (53 indexed by SCI and 93 by EI); co-authored 3 book chapters and filed 9 granted patents. He is a member of SPIE.